**Enhanced electronic correlations and altermagnetic ground state of two-dimensional $CsCr_3Sb_5$ monolayers**

Z. H. Guan[1], Z. L. Peng[1], W. Z. Zhuo[2,*], G. Tian[1], Z. P. Hou[1], D. Y. Chen[1], Z. Fan[1], X. B. Lu[1], X. S. Gao[1], M. H. Qin[1,†], and J. –M. Liu[1,3]

[1]*Guangdong Provincial Key Laboratory of Quantum Engineering and Quantum Materials and Institute for Advanced Materials, South China Academy of Advanced Optoelectronics, South China Normal University, Guangzhou 510006, China*

[2]*School of Optoelectronic Engineering, Guangdong Polytechnic Normal University, Guangzhou 510665, China*

[3]*Laboratory of Solid State Microstructures, Nanjing University, Nanjing 210093, China*

*wzzhuo@gpnu.edu.cn

†qinmh@scnu.edu.cn

**[Abstract]** Recently, layered corrected kagome metal $CsCr_3Sb_5$ have garnered significant attention attributed to its flat bands near the Fermi level ($E_F$) and altermagnetic ground state [ Yi Liu *et al*., Nature 632, 1032 (2024)]. However, the van Hove singularities (vHSs) in bulk $CsCr_3Sb_5$ are far away from the $E_F$, while an effective modulation of VHS toward the $E_F$ is essential for exploring intriguing electron transport properties. In this work, using first-principles calculations, we investigate electronic structures of two-dimensional (2D) $Cr_3Sb_5$ and $CsCr_3Sb_5$ monolayers which may be mechanically exfoliated from bulk materials. Notably, it is revealed that both flat bands and vHSs simultaneously reside in close proximity to the $E_F$ in $Cr_3Sb_5$ monolayer, signifying enhanced electronic correlations. Importantly, a tensile strain further shifts the incipient flat bands and vHSs of two monolayers simultaneously to the vicinity of the $E_F$, suggesting strain tunable electronic correlations and concomitant quantum effects. Furthermore, altermagnetic ground state is also revealed due to retained mirror symmetry between two sublattices in these two monolayers. Thus, this work advances understanding and modulations of electronic properties of 2D $CsCr_3Sb_5$ monolayers, strengthening their great potential for exploring unconventional quantum phenomena and altermagnetism.

Keywords: kagome metals, flat band, van Hove singularities, electronic correlation, altermagnetism

## I. INTRODUCTION

Kagome materials with corner-sharing triangular networks have attracted significant research interest due to their unique electronic structures including flat bands, Dirac fermions, and van Hove singularities (vHSs) owing to the geometrical frustration [1,2]. Specifically, fractional quantum Hall effect and quantum anomalous Hall effect may emerge when the flat band is near to the Fermi level ($E_F$) [1,3-4]. Furthermore, vHSs and the Dirac fermions around $E_F$ can induce unconventional superconductivity and/or other novel electron transport properties [5-7]. Therefore, kagome materials provide a promising platform for exploring magnetic frustration, electronic correlation, and quantum effects [8-13].

Among these abundant kagome materials, vanadium-based kagome metals $AV_3Sb_5$ (A = K, Rb, and Cs) draw particular attention due to their intriguing emergent properties [14,15], including unconventional superconductivity [15-20], charge density waves (CDW) [15,21-25], potential Majorana zero modes [16], and edge supercurrent [19]. The impetus of these unconventional properties is attributed to the combination of Coulomb interactions and vHSs near the $E_F$ [26,27]. In $AV_3Sb_5$, however, the flat band lies far away from the $E_F$, and the materials are weakly electron correlated without intrinsic magnetism [28,29].

Compared to $AV_3Sb_5$, compound $CsCr_3Sb_5$ exhibits an incipient flat band near the $E_F$ [30-34]. More importantly, at ~ 55K, the system undergoes a phase transition to an antiferromagnetic phase which is suggested to be an altermagnetic spin density wave (SDW) state [30-31,35]. It is well noted that altermagnetic order is characterized by compensated antiferromagnetic spin polarization and momentum-dependent band splitting, which holds great application potential in future high-density and low-consuming storage devices [36-37]. However, both the vHSs and Dirac cones are far away from the $E_F$, strongly preventing these fascinating electron transport properties. Along this line, it is meaningful to tune the vHSs in $CsCr_3Sb_5$ to the vicinity of $E_F$. However, the energetic positions of vHSs and flat bands are strongly fixed and resistant to be tuned in bulk $CsCr_3Sb_5$, while two-dimensional (2D) materials serve as a potential alternative [31].

To be specific, 2D materials impose strong confinement on out-of-plane electron motion owing to low-dimensional physical properties, resulting in pronounced modifications in electronic density of states (DOS) and band structures upon dimensional reduction [38-41]. Moreover, their atomically thin nature eliminates all channels in bulk for strain relaxation—such as dislocations and interlayer sliding—thereby ensuring that the applied strain is fully retained within 2D plane and directly reshapes the electronic wavefunctions [42]. As a result, electronic structures of 2D materials are highly susceptible to applied strain. As a matter of fact, 2D $AV_3Sb_5$ (A = K, Rb, and Cs) has been successfully realized in experiments via mechanical exfoliation [43,44], and the rearrangement of vHSs caused by dimensional reduction are theoretically substantiated [45]. Furthermore, recent experimental observations have uncovered prominent compressive and tensile strain effects on the vHSs in $CsV_3Sb_5$ [46], suggesting that the electronic structure of 2D $CsCr_3Sb_5$ is susceptible to efficient strain engineering. Furthermore, 2D $CsCr_3Sb_5$ are more readily integrated into practical devices, compared to bulk material. However, electronic structures and associated tunability of 2D $CsCr_3Sb_5$—characterized by flat bands, vHSs, and magnetic ground state—have remained elusive to date, as far as we know.

In this work, using first-principles calculations, we study electronic structures of 2D $Cr_3Sb_5$ and $CsCr_3Sb_5$ monolayers that are most likely to be realized in experiments. It is revealed that both the incipient flat band and vHSs appear near the $E_F$ in $Cr_3Sb_5$ monolayer, suggesting enhanced electronic correlations by dimension reduction. Notably, a tensile strain further shifts the incipient flat bands and vHSs of two monolayers simultaneously to the vicinity of the $E_F$, thus dramatically enhance their electronic corrections. Furthermore, both the two monolayers also exhibit altermagnetism due to the retained mirror symmetry between two sublattices.

## II. COMPUTATIONAL METHODS

We conduct the first-principles calculations using the Vienna Ab initio simulation package (VASP) based on the density functional theory (DFT) [47,48]. In all calculations, the Perdew-Burke-Ernzerhof for solids (PBEsol) [49] form approximation is used, and spin-orbit coupling is not included, following earlier work. The van der Waals correction is included within the zero damping DFT-D3 method of Grimme [50]. The kinetic energy cutoff for the plane wave basis is 450 eV, and the force criteria for optimizing the structures is 0.01 eV/Å. The $CsCr_3Sb_5$ monolayers is simulated using a periodic supercell with a vacuum spacing of more than 20 Å. A biaxial strain $\frac{a-a_0}{a_0} \times 100\%$ ($a_0$ and $a$ represent the initial and strained lattice constants, respectively.) is applied to adjust the lattice and manipulate electronic structures.

## III. RESULTS AND DISCUSSION

### A. Lattice structures and electronic structures of 2D $CsCr_3Sb_5$ monolayers

$CsCr_3Sb_5$ is isostructural to $CsV_3Sb_5$, and its high-temperature bulk structure adopts the undistorted P6/mmm space group in which Cr atoms form a kagome lattice through corner-sharing triangles (Fig. 1a). Considering the successful mechanical exfoliation of monolayer $CsV_3Sb_5$ [43,44], isolation of monolayer $CsCr_3Sb_5$ is highly accessible considering their structural similarity. In experiments, mechanical thinning is found to proceed via preferential cleavage at the weakly bonded Cs–Sb interfaces [43]. Motivated by this property, we constructed a freestanding $Cr_3Sb_5$ monolayer (Fig. 1b) and considered six possible configurations of monolayer $CsCr_3Sb_5$ (Fig. 1c) within a $2 \times 2$ supercell (Fig. S1). Energetic analysis indicates that the $Cr_3Sb_5$ monolayer is with a rather low energy, and the B-type $CsCr_3Sb_5$ monolayer holds the lowest energy among these six $CsCr_3Sb_5$ configurations (Table I). Consequently, we focus our attention on the $Cr_3Sb_5$ and B-type $CsCr_3Sb_5$ monolayers, while briefly discuss other configurations in the Supplementary Information. Notably, upon dimensional reduction, the B-type $CsCr_3Sb_5$ monolayer undergoes a lattice distortion into a rectangular geometry with a $\sqrt{3} \times 1$ translational symmetry [44], resulting in a concomitant folding of the Brillouin zone (Fig. 1d).

The electronic structure of nonmagnetic bulk $CsCr_3Sb_5$ is shown in Fig. 2a which is well consistent with the earlier report [30,31]. Specifically, a long flat band originating from the $d_{xz}/d_{yz}$ orbitals is located around 300 meV above the $E_F$, and flat bands derived from the $d_{z^2}$ orbital lie in the vicinity of the $E_F$. Both vHSs and Dirac cones are rather far away from the $E_F$. Unlike $AV_3Sb_5$, additional vHSs are identified around the K point at approximately 280 meV below the $E_F$ [31], exhibiting a rather flat dispersion along K ↔ M (black-boxed areas in Figs. 2a).

In bulk $CsCr_3Sb_5$, adjacent layers provide additional degrees of freedom for electron motion. The dimensional reduction to 2D limit dramatically enhances the effective Coulomb repulsion, which drives a pronounced upward shift of the electronic bands [51,52], as shown in Figs. 2b and 2c which depict the projected bands of B-type $CsCr_3Sb_5$ and $Cr_3Sb_5$ monolayers, respectively. Consequently, both the flat band and the vHSs shift upward in these two monolayers. Fig. 2d illustrates the evolution of the Cr *d*-orbital band near the $E_F$ upon dimensional reduction from three-dimensional to 2D. In 2D $CsCr_3Sb_5$ monolayer, Brillouin zone folding is analogous to that in $AV_3Sb_5$ monolayers, which induces a rearrangement of vHSs approximately 400 meV below the $E_F$ (blue box in Fig. 2b). Furthermore, the repositioning of vHS at the Γ point is combined with the flat band dispersion along M ↔ K path, giving rise to an extending flat band across Γ ↔ K ↔ M at 200 meV below the $E_F$. Furthermore, the removal of Cs layer further amplifies the Coulomb interaction, resulting in a substantial upward shift of the incipient flat band and vHSs in $Cr_3Sb_5$ monolayer. As a result, both the vHSs and Dirac cones get close to the $E_F$ (Fig. 2c). It is noted that the vHSs near the $E_F$ in $AV_3Sb_5$ strongly boost electron correlations, and similar phenomenon is expected in $Cr_3Sb_5$ monolayer. Thus, exotic quantum states such as charge density waves and anomalous Hall effect could be induced.

### B. Strain modulated electronic structures of 2D $CsCr_3Sb_5$ monolayers

2D $CsCr_3Sb_5$ and $Cr_3Sb_5$ monolayers also provide platforms for investigating strain-tunability of electronic structures, noting that effective modulation of the CDW order and vHSs by strain in bulk $CsV_3Sb_5$ have been reported [46]. In this section, we systematically explore the evolution of these features in 2D $Cr_3Sb_5$ and $CsCr_3Sb_5$ monolayers under applied strain, and give their projected bands under various strains in Figs. 3 and 4, respectively. The results clearly show that the flat bands and vHSs of two monolayers are highly susceptible to strain. Generally, a compressive strain drives upward shift of $d_{xz}/d_{yz}$-derived bands and downward shift of $d_{z^2}$-derived bands, and a tensile strain induces inverse shifts as depicted in Figs. 3g and 4g.

In $Cr_3Sb_5$ monolayer, the compressive strain drives an upward shift of the incipient flat band, and induces an additional vHS with a short flat band (black boxes in Figs. 3a and 3c) derived from the $d_{xz}/d_{yz}$ orbital along M ↔ K path. Consequently, under −2% strain (Fig. 3a), the additional vHS with the short flat band is tuned to align almost with the $E_F$, generating a sharp peak in the DOS. Conversely, the $d_{z^2}$-derived bands shift downward under the compressive strain, and the short flat band along the M ↔ K path (orange box) is driven down to the $E_F$ under −6% strain, inducing a distinct $d_{z^2}$-derived DOS peak (Fig. 3c). On the other hand, under the tensile strains, the incipient flat band arising from the $d_{xz}/d_{yz}$ orbitals gradually shift downward to the $E_F$, while the broadening of the $d_{z^2}$ bands decrease and the band shift upward. As a result, the vHS induced by the combined contributions of the $d_{z^2}$, $d_{x^2-y^2}$ and $d_{xy}$ orbitals (blue box) is driven to the vicinity of the $E_F$ by the tensile strain.

The strain-dependent evolution of the electronic structures in B-type $CsCr_3Sb_5$ monolayer is similar to that of $Cr_3Sb_5$. Under a compressive strain, the $d_{xz}/d_{yz}$-derived bands shift upward progressively, while the $d_{z^2}$-derived bands shift downward, tuning the short flat bands at the K point directly to the $E_F$ (orange boxes in Figs. 4a-4c). Notably, a 6% tensile strain shifts the incipient flat band into immediate vicinity of the $E_F$, and the corresponding DOS peak persists (Fig. 4f). Furthermore, the vHSs (blue box), which are originated from three orbitals and rearranged via Brillouin zone folding, also shifts upward toward the $E_F$. In addition, the width of $d_{z^2}$-derived band contracts markedly, and the bands gradually become flat with the increasing tension. Under a 5% tensile strain, for example, a prominent long-range flat band dominated by the $d_{z^2}$ orbital emerges at approximately 1 eV, extending across the entire Brillouin zone (Fig. 4e). These collective band-structure modifications under tensile strain signify a pronounced enhancement of electronic correlations, suggesting the potential of unconventional electronic transport properties.

The strain-modulated electronic structures in these monolayers mainly originate from lattice deformations. Under a tensile strain, in-plane lattice expansion and concomitant out-of-plane contraction enhance the localization of electrons within the $d_{z^2}$ orbital, thereby amplifying the effective Coulomb repulsion. The enhanced electronic localization and Coulomb repulsion lead to a narrowing of the bandwidth and an upward shift of $d_{z^2}$-derived bands (highlighted by the orange boxes in Fig. S5). Consequently, the vHS arising from the $d_{z^2}/d_{x^2-y^2}/d_{\mathrm{xy}}$ orbitals shift upward toward the $E_{\mathrm{F}}$. In contrast, the Coulomb repulsions acting on the $d_{xz}/d_{yz}$ orbitals are suppressed, causing the flat band to descend toward the $E_{\mathrm{F}}$. Collectively, these band-structure modifications demonstrate an enhancement of electronic correlations under the tensile strain. Conversely, under the compressive strain, the $d_{xz}/d_{xz}$ -derived incipient flat band shifts upward, while the $d_{z^2}$ -derived bands shift downward. Generally, the evolutions of energy bands of Cr orbitals near the $E_{\mathrm{F}}$ under strains in $Cr_3Sb_5$ and B-type $CsCr_3Sb_5$ monolayers are summarized in Figs. 3g and 4g, respectively.

It is noted that strongly correlated phenomena in bulk $CsCr_3Sb_5$ are attributed to the incipient flat band derived from the $d_{xz}/d_{xz}$ orbitals [31-33]. The incipient flat band can be tuned either toward or away from the $E_{\mathrm{F}}$ via strain engineering, providing a viable pathway to manipulate these emergent quantum states. Under the tensile strain, the incipient flat band and the vHS are driven toward the $E_{\mathrm{F}}$, resulting in an enhancement of electronic correlations in both $CsCr_3Sb_5$ and $Cr_3Sb_5$ monolayers. Under the compressive strain, the short flat bands arising from the $d_{z^2}$ and $d_{xz}/d_{xz}$ orbitals are tuned to the vicinity of the $E_{\mathrm{F}}$, while the incipient flat bands recede from the $E_{\mathrm{F}}$. Such a selective tuning may enable investigations on respective effects of these short flat bands. As a result, the strain-induced evolution of the flat bands and vHSs suggests a high degree of tunability inherent to intriguing quantum phenomena.

**C. Possible altermagnetic ground state in 2D $CsCr_3Sb_5$ monolayers**

In experiments, a stripe-like structural modulation emerges in bulk $CsCr_3Sb_5$ below 55 K, coinciding with a concurrent structural distortion and a SDW phase transition [30]. Importantly, recent high-throughput calculations and experiments have suggested this SDW phase to be an altermagnetic ground state (Fig. 5a) [31,35]. Notably, the altermagnetic order may also persist in 2D $Cr_3Sb_5$ and $CsCr_3Sb_5$ monolayers. To be specific, this order still possesses the lowest energy in these typical magnetic states (see Fig. S11 and Table III), suggesting that the altermagnetic state still is the ground state in 2D $CsCr_3Sb_5$ monolayers.

The spin-resolved band structures of $Cr_3Sb_5$ and $CsCr_3Sb_5$ monolayers are presented in Fig. 5c and Fig. 5d, respectively. For $Cr_3Sb_5$ monolayer, the crystal symmetry remains unchanged compared to bulk material. Following the SDW modulation, the altermagnetic configuration (Fig. S11) is still identified as the ground state in $Cr_3Sb_5$ monolayer with two sublattices connected by a $\{M_{yz}|(0,1/2,0)\}$ or a $\{M_{xz}|(1/2,0,0)\}$ symmetry operation (shown in Fig. 5b). For B-type $CsCr_3Sb_5$ monolayer with the altermagnetic configuration, the space group is reduced to No. 32, Pba2. Compared with bulk system, $CsCr_3Sb_5$ monolayer has a lower symmetry along the *z*-direction due to the absence of the capping Cs layer. Importantly, the symmetry operation connecting two sublattices is preserved, ensuring the altermagnetic ground state (Fig. 5d). Furthermore, a specific $d_{z^2}$-derived band (blue circles in Figs. S13 and S14) undergoes a bandwidth reduction and moves to the vicinity of the $E_F$ under −4% strain, resulting in a very sharp peak in DOS. These results reveal that the electronic structures in $CsCr_3Sb_5$ monolayers are highly tunable even when the SDW is taken into account.

### D. Brief Discussion

So far, we have clarified the important roles of dimension and strain in modulating the electronic structures of 2D $CsCr_3Sb_5$, which provides valuable insights for exploring electronic correlation and intriguing quantum effects. For example, the incipient flat band, vHSs and Dirac cones can be tuned close to the $E_F$, suggesting enhanced electronic correlations and unconventional transport properties. Moreover, the short flat bands arising from different orbitals can be tuned toward or away from the $E_F$ under various strains, enabling the investigation on respective effects of these specific short flat bands. Importantly, the altermagnetic ground state preserves in 2D $CsCr_3Sb_5$ monolayers, attributing to the symmetry operation of $\{M_{yz}|(0,1/2,0)\}$ or a $\{M_{xz}|(1/2,0,0)\}$ between two sublattices. As a result, the coexistence of altermagnetism and enhanced electronic correlations in 2D kagome monolayers implies their unique potentials in future spintronic device design.

In experiments, 2D $CsV_3Sb_5$ has been realized using an exfoliation scheme, and the same method may be transferred to design 2D $CsCr_3Sb_5$ monolayers, considering the lattice similarity of two materials. In this work, the simultaneous appearance of the incipient flat bands, vHSs and Dirac cones near the $E_F$ in the $CsCr_3Sb_5$ monolayers implies that it may concurrently host unconventional properties, such as charge density waves, anomalous Hall effect, and other transport-related characteristics. Furthermore, the strain-induced evolution of flat bands and vHSs in 2D $CsCr_3Sb_5$ indicates a pronounced tunability of these quantum phenomena, which highly deserves to be checked in future experiments and theoretical calculations.

## IV. CONCLUSIONS

In conclusion, we have studied the electronic structures and magnetic state of 2D $CsCr_3Sb_5$ monolayers using the first-principles calculations. Both flat bands and vHSs simultaneously are rather close to the $E_F$ in $Cr_3Sb_5$ monolayer. Interestingly, the incipient flat bands and vHSs are highly susceptible to strain engineering, and tensile strain shifts the incipient flat bands and vHSs to the vicinity of the $E_F$, significantly enhancing electronic correlations. Furthermore, both $Cr_3Sb_5$ and B-type $CsCr_3Sb_5$ monolayers exhibit the altermagnetic ground state, demonstrating new candidates to realize 2D altermagnetism.

## ACKNOWLEDGMENTS

The work is supported by the Natural Science Foundation of China (Grants No.

U22A20117, and No. 52371243), the Guangdong Provincial Quantum Science strategic initiative (Grant No. GDZX2401002), the Guangdong Basic and Applied Basic Research Foundation (Grant No. 2024A1515012665, No. 2023B1515020112, and No. 2024B1515020076).

*Phys. Rev. B* **2019**, *99*, 245114.

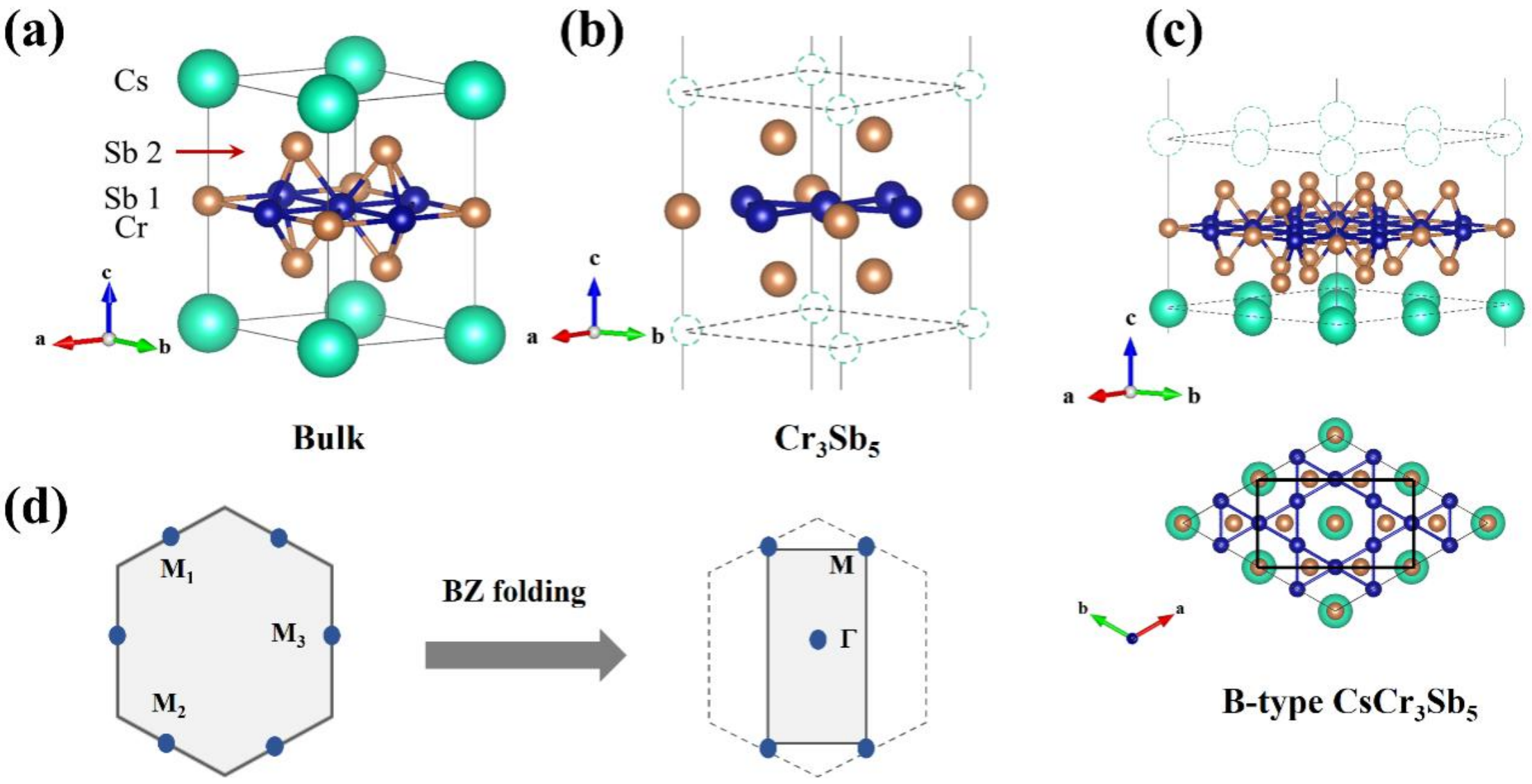

**FIG. 1.** Crystal structures for (a) High-temperature crystal structure of bulk $CsCr_3Sb_5$, (b) 2D $Cr_3Sb_5$ monolayer, and (c) B-type $CsCr_3Sb_5$ monolayer. (d) Schematic illustration of Brillouin zone (BZ) folding from bulk material to $Cr_3Sb_5$ monolayer (left) and the vHSs in momentum space rearranged in B-type $CsCr_3Sb_5$ monolayer (right).

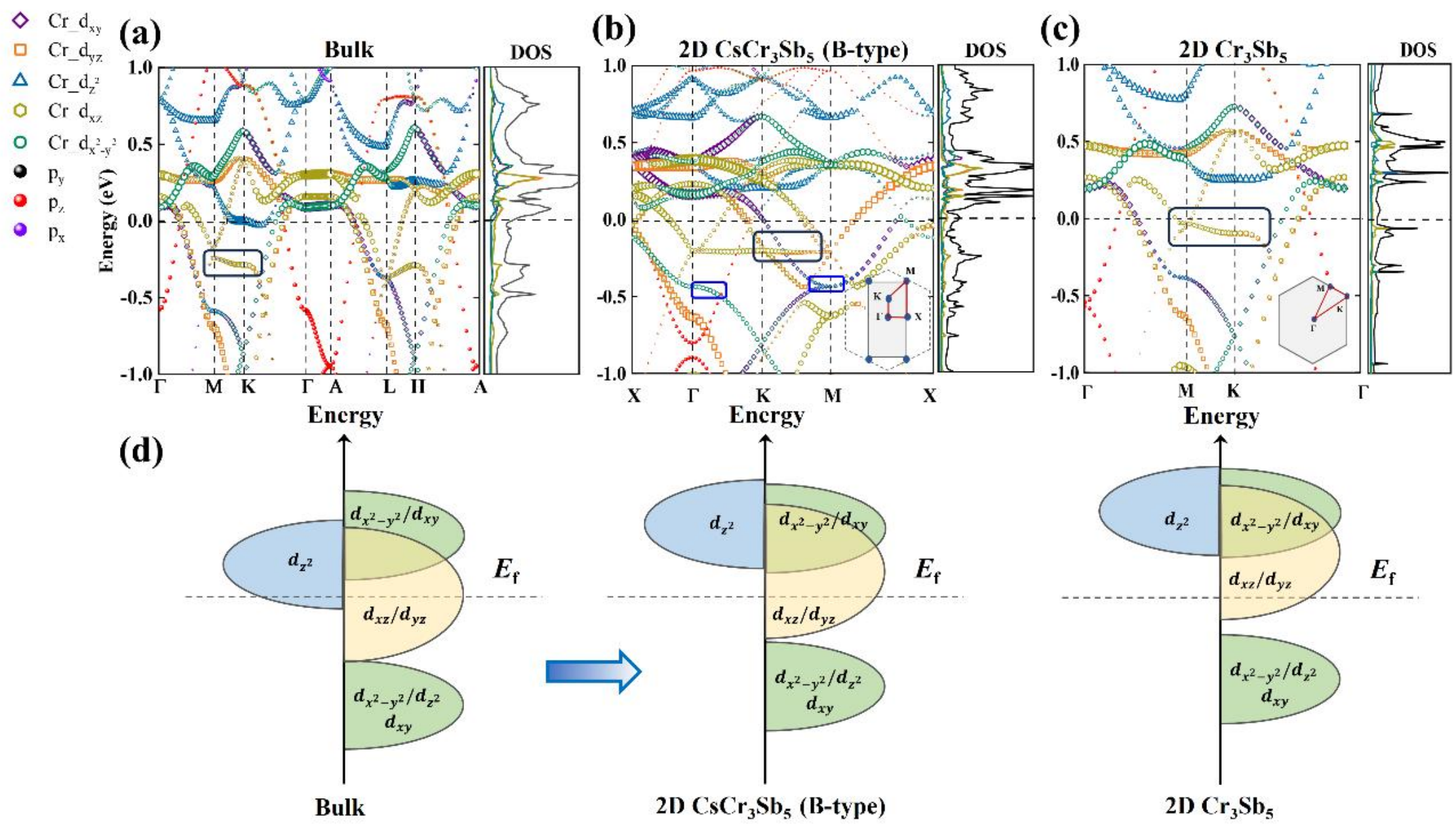

**FIG. 2.** Projected bands for (a) bulk $CsCr_3Sb_5$, (b) 2D B-type $CsCr_3Sb_5$ monolayer, and (c) $Cr_3Sb_5$ monolayer. (d) Schematic of the Cr *d*-orbitals bands near the $E_F$, showing modifications of electronic structures from three dimensional to 2D. The boxed regions indicate the vHSs. The black box marks an extra vHS and a short flat band along K ↔ M in $CsCr_3Sb_5$, and the blue box indicates vHSs rearrangement.

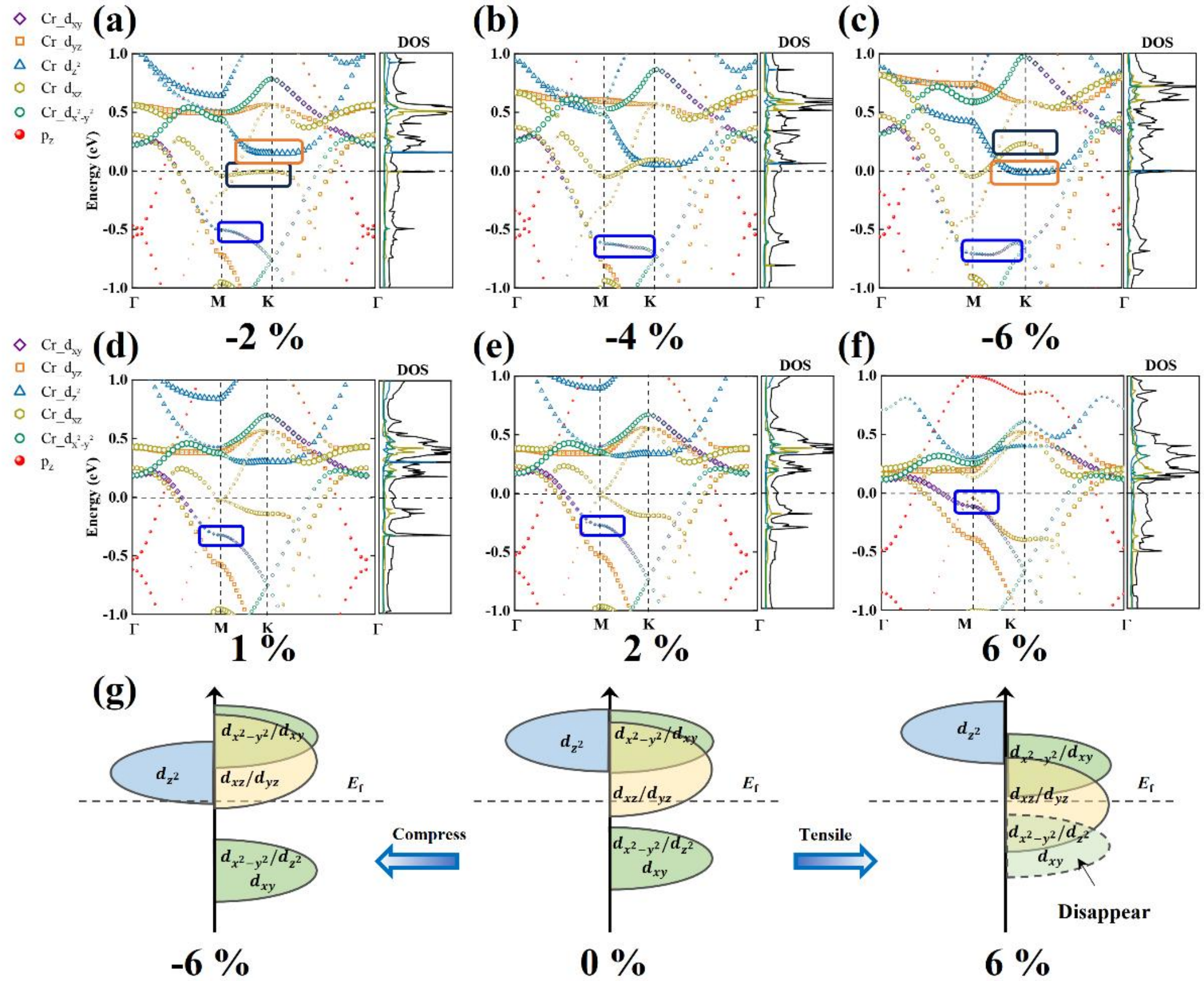

**FIG.3.** Projected bands of 2D $Cr_3Sb_5$ monolayer under the compressive strain of (a) −2%, (b) −4%, and (c) −6%, and tensile strain of (d) 1%, (e) 2%, (f) 6%. (g) Schematic of the Cr *d*-orbitals bands near the $E_F$, showing strain-induced modifications of electronic structures in 2D $Cr_3Sb_5$ monolayer. the green (orange) boxed areas indicate the short flat bands originating from the $d_{z^2}$ ($d_{xz}/d_{yz}$) that gradually shift downward (upward) toward the $E_F$ under compressive strain. The blue boxes highlight the positions of vHS originating from the $d_{x^2-y^2}/d_{z^2}/d_{xy}$ orbitals under various strains.

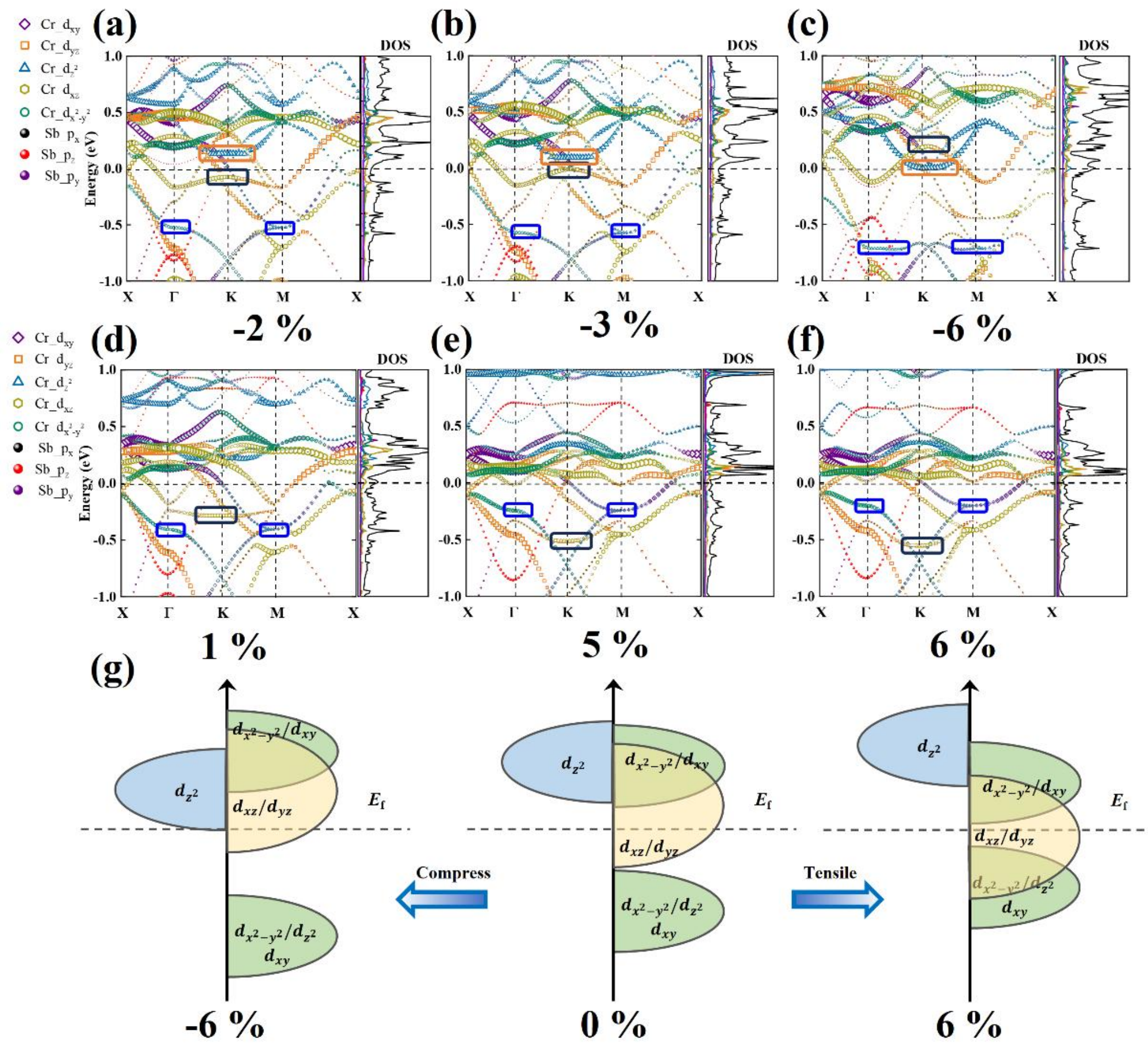

**FIG.4.** Projected bands of 2D B-type $CsCr_3Sb_5$ monolayer under the compressive strain of (a) −2%, (b) −3%, (c) −6%, and tensile strain of (d) 1%, (e) 5%, (f) 6%. (g) Schematic of the Cr *d*-orbitals band near the $E_F$, showing strain-induced modifications of the electronic structures in 2D B-type $CsCr_3Sb_5$ monolayer. The boxed region here is interpreted in the same way as in Figure 3.

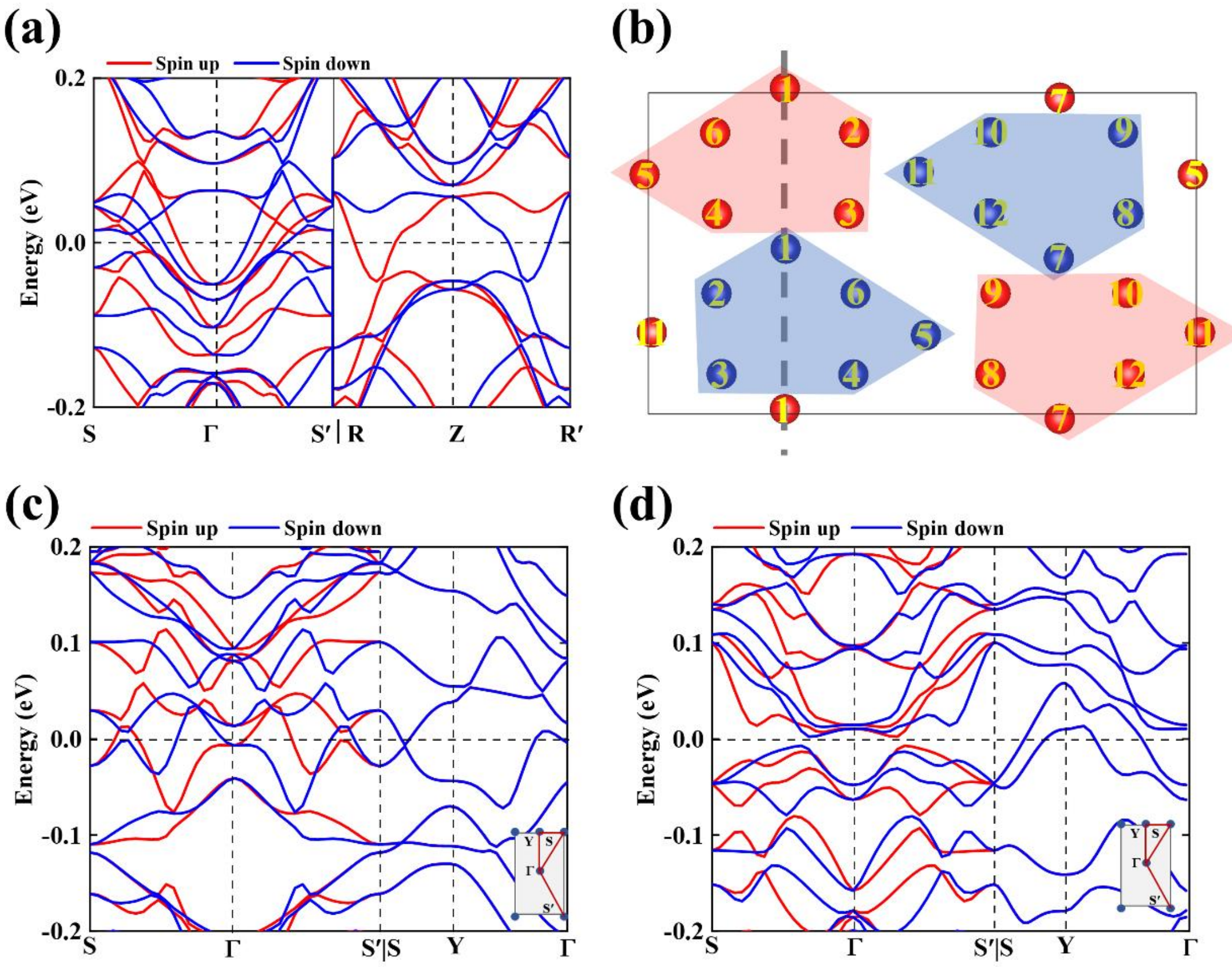

**FIG.5.** Band structures of the $4 \times \sqrt{3}$ (a) bulk $CsCr_3Sb_5$, (c) $Cr_3Sb_5$ monolayer, and (d) B-type $CsCr_3Sb_5$ monolayer. (b) Ground state SDW pattern (failed AF-SOD). The spin-up and spin-down sublattices are marked in red and blue with symmetry-related Cr sites of opposite spin labeled by the same index, respectively.

**Supplementary materials for "Enhanced electronic correlations and altermagnetic ground state of two-dimensional $CsCr_3Sb_5$ monolayers"**

Z. H. Guan[1], Z. L. Peng[1], W. Z. Zhuo[2,*], G. Tian[1], Z. P. Hou[1], D. Y. Chen[1], Z. Fan[1], X. B. Lu[1], X. S. Gao[1], M. H. Qin[1,†], and J. –M. Liu[1,3]

[1]*Guangdong Provincial Key Laboratory of Quantum Engineering and Quantum Materials and Institute for Advanced Materials, South China Academy of Advanced Optoelectronics, South China Normal University, Guangzhou 510006, China*

[2]*School of Optoelectronic Engineering, Guangdong Polytechnic Normal University, Guangzhou 510665, China*

[3]*Laboratory of Solid State Microstructures, Nanjing University, Nanjing 210093, China*

## A. Other possible Lattice structures and electronic structures of $CsCr_3Sb_5$ monolayers.

Unlike bulk $CsCr_3Sb_5$ in which Cs atoms are shared with adjacent two unit cells, Cs atoms in $CsCr_3Sb_5$ monolayer belong to one unit cell. To preserve stoichiometry, we constructed $CsCr_3Sb_5$ monolayers using 2 × 2 supercells with half of Cs atoms are removed. Considering distinct Cs arrangements within the supercell, six monolayer configurations with the same stoichiometry as that of bulk system are constructed (Fig. 1c, Fig. S1, and Fig. S2(a)), in addition to 2D $Cr_3Sb_5$ monolayer shown in Fig. 1b.

The energies of these structures are given in Table I. Among these six structures with the same stoichiometry, the B-type $CsCr_3Sb_5$ monolayer has the lowest energy, which is slightly lower that the S-type $CsCr_3Sb_5$ monolayer (Fig. S2a). Similarly, in the S-type $CsCr_3Sb_5$ monolayer, a rearrangement of van Hove singularities (vHSs) is also observed (around 400 meV below the $E_F$ shown in the blue-boxed region in Fig. S2b), resulting from the folding of the Brillouin zone.

* wzzhuo@gpnu.edu.cn

† qinmh@scnu.edu.cn

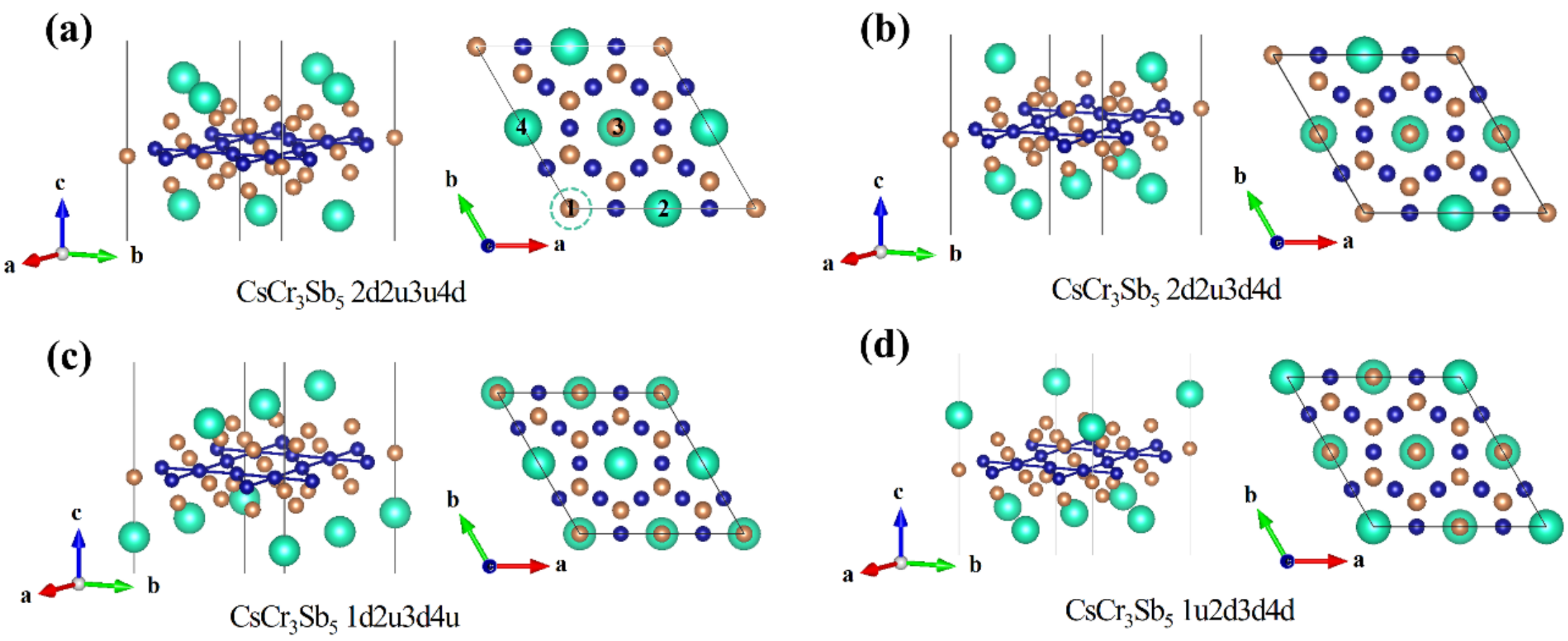

FIG. S1. Four monolayer structures of $CsCr_3Sb_5$. Possible atomic sites (green dashed open circles; labeled as 1 to 4 with up and down positions) of Cs atoms in the monolayer structures.

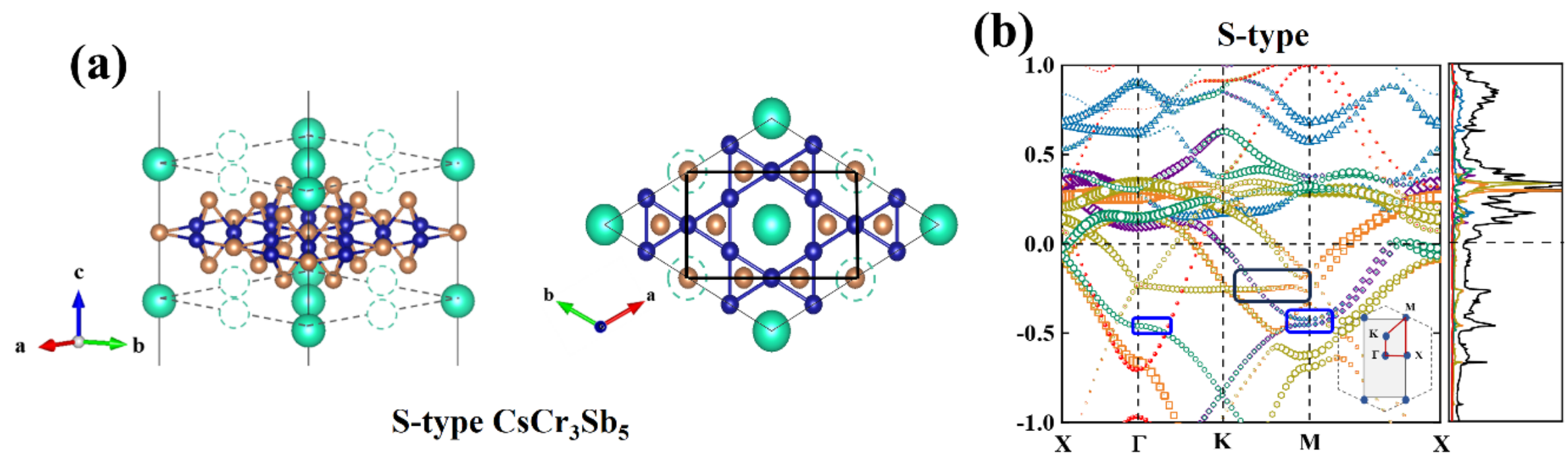

FIG. S2. (a) Crystal structures and (b) projected bands of the S-type $CsCr_3Sb_5$ monolayer. The black box marks an extra vHS and a flat band along K ↔ M in $CsCr_3Sb_5$, and the blue boxes indicates vHSs rearrangement.

Table I. Total energies (in meV/f.u.) of seven types of $CsCr_3Sb_5$ monolayers.

| Configuration | 2D $Cr_3Sb_5$ | B-type | S-type | 2d2u3u4d | 2d2u3d4d | 1d2u3d4u | 1u2d3d4d |
|---|---|---|---|---|---|---|---|
| Energy (meV/f.u.) | -405.9 | -273.6 | -210.2 | -209.2 | -206.4 | -207.8 | -206.8 |

## B. Strain modulated electronic structures of 2D $CsCr_3Sb_5$ monolayers

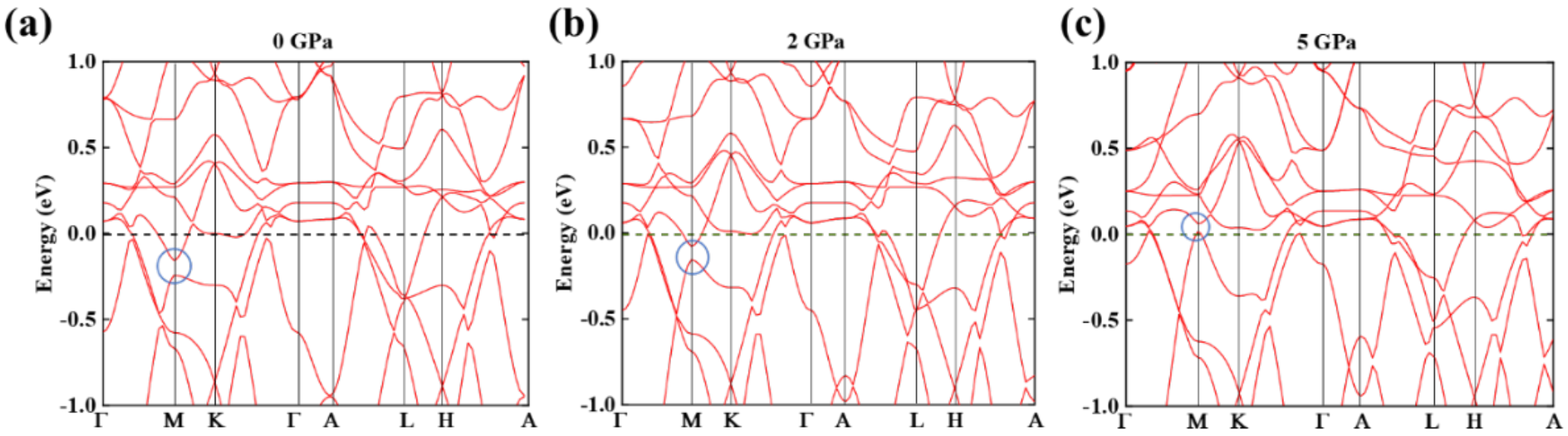

FIG. S3. Electronic band structures of bulk $CsCr_3Sb_5$ under strain of (a) 0 GPa, (b) 2 GPa, and (c) 5 GPa. The blue circles show the Lifshitz transition under pressure, while the incipient flat band and vHSs hardly be affected.

In bulk $CsCr_3Sb_5$, both the vHSs and initial flat bands remain nearly unchanged under various compressive strain, as shown in Fig. S3, revealing their robustness against strain. Strong tunability of electronic structures in $CsCr_3Sb_5$ monolayers by applied strain has been revealed, and the detailed results are presented here. Under tensile strain, the electronic localization and Coulomb repulsion are enhanced, and lead to a narrowing of the bandwidth and an upward shift of $d_{z^2}$-derived bands, showing a prominent long-range flat band at approximately 1 eV (highlighted by the orange boxes in Fig. S4). Furthermore, Figs. S5-S8 present bands modulated by strain in the $Cr_3Sb_5$ and B-type $CsCr_3Sb_5$ monolayers, and Fig. S9 is the schematic of the electron distribution for different Cr-sublattices.

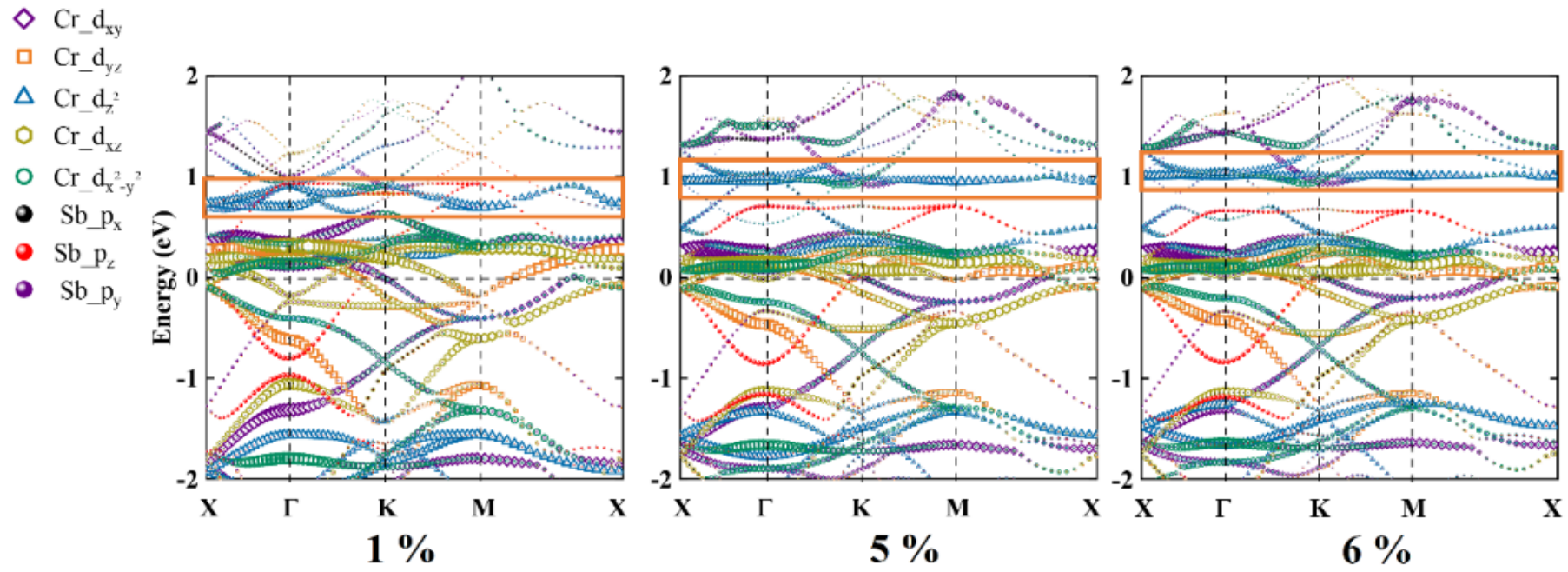

FIG. S4. Projected bands with wide range for 2D $CsCr_3Sb_5$ under various strains. The orange boxes indicate the upward shift of the $d_{z^2}$-derived bands.

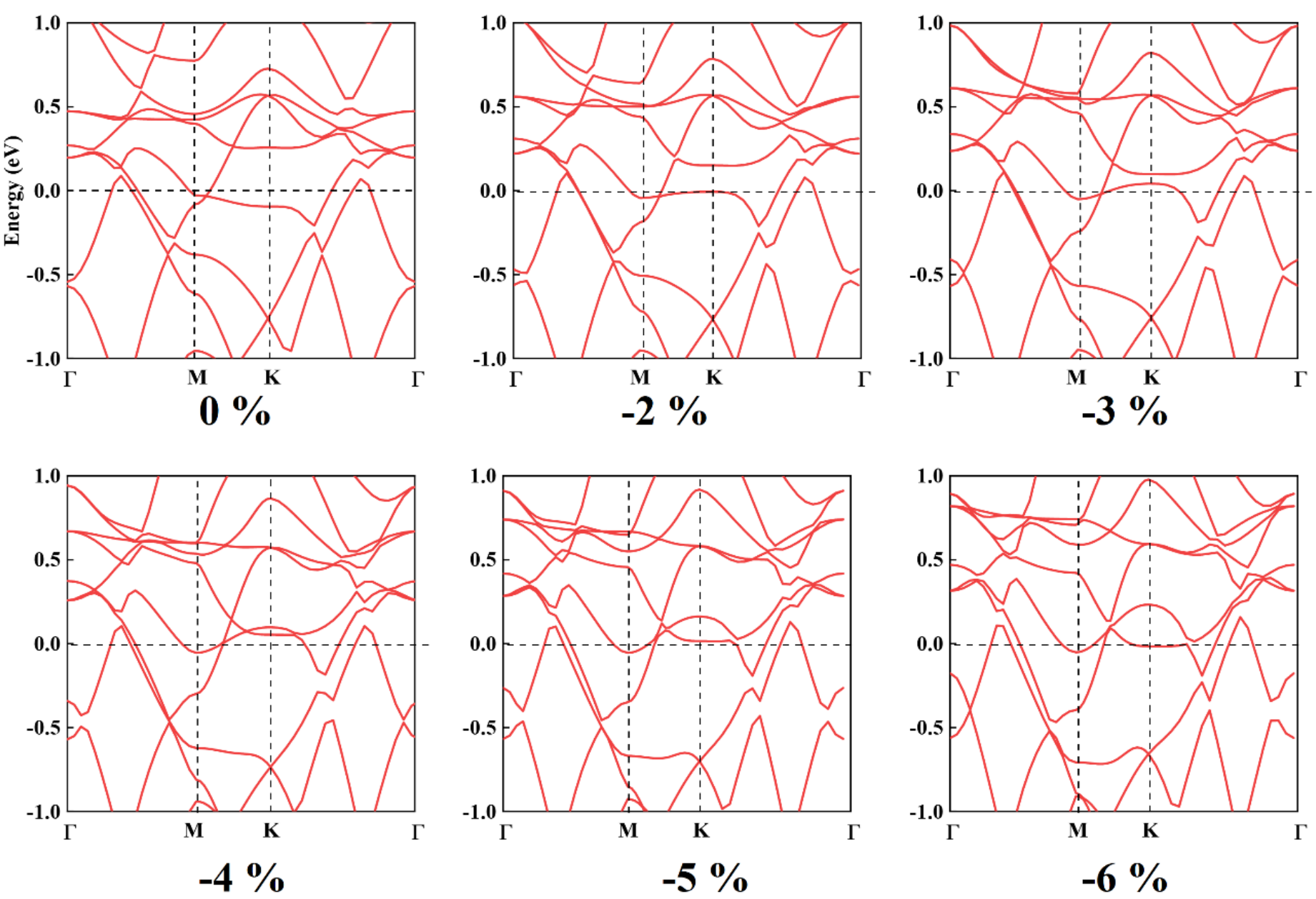

FIG. S5. Band structures of $Cr_3Sb_5$ monolayer under various compressive strains.

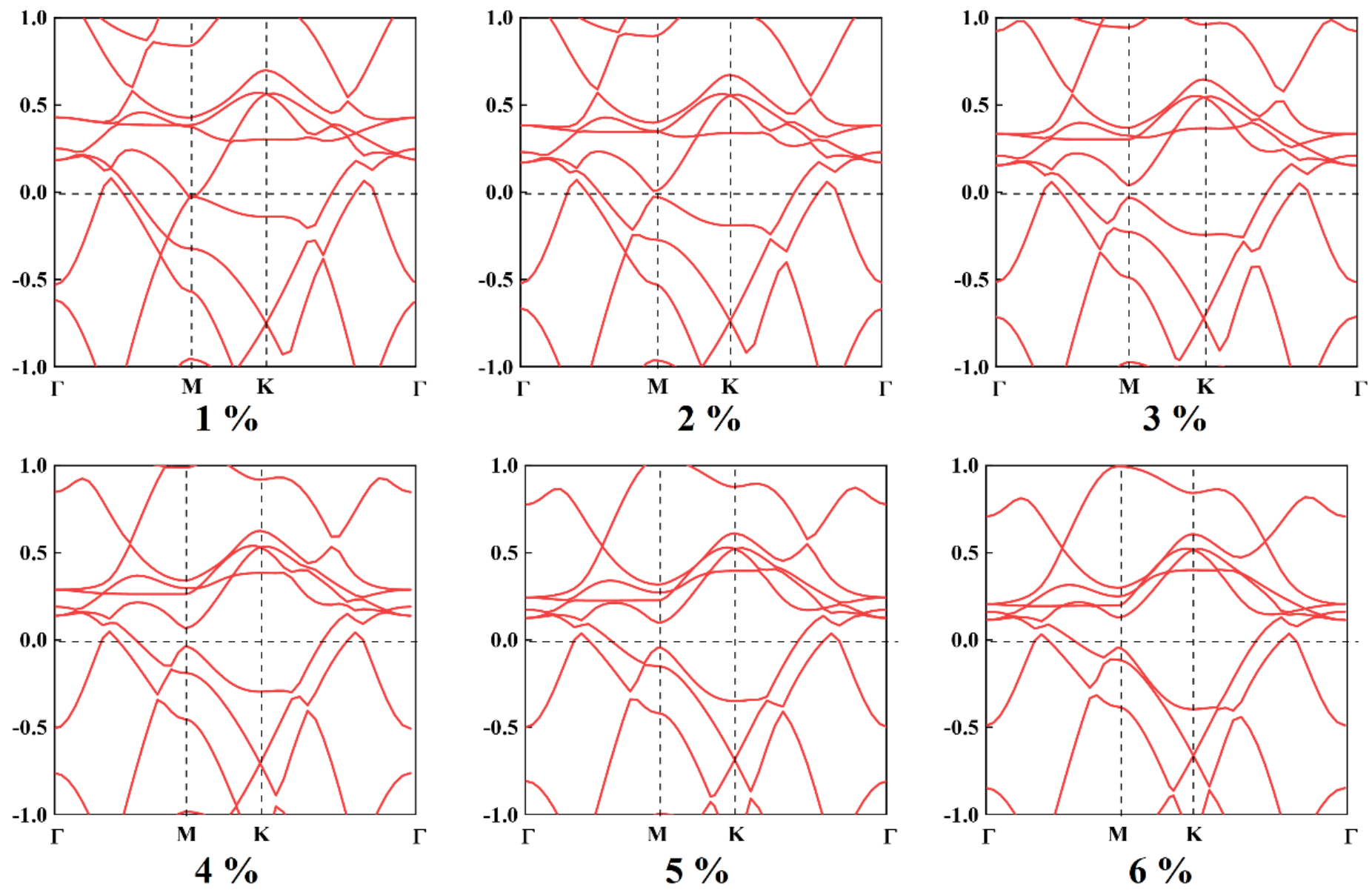

FIG. S6. Band structures of 2D $Cr_3Sb_5$ monolayer under various tensile strains.

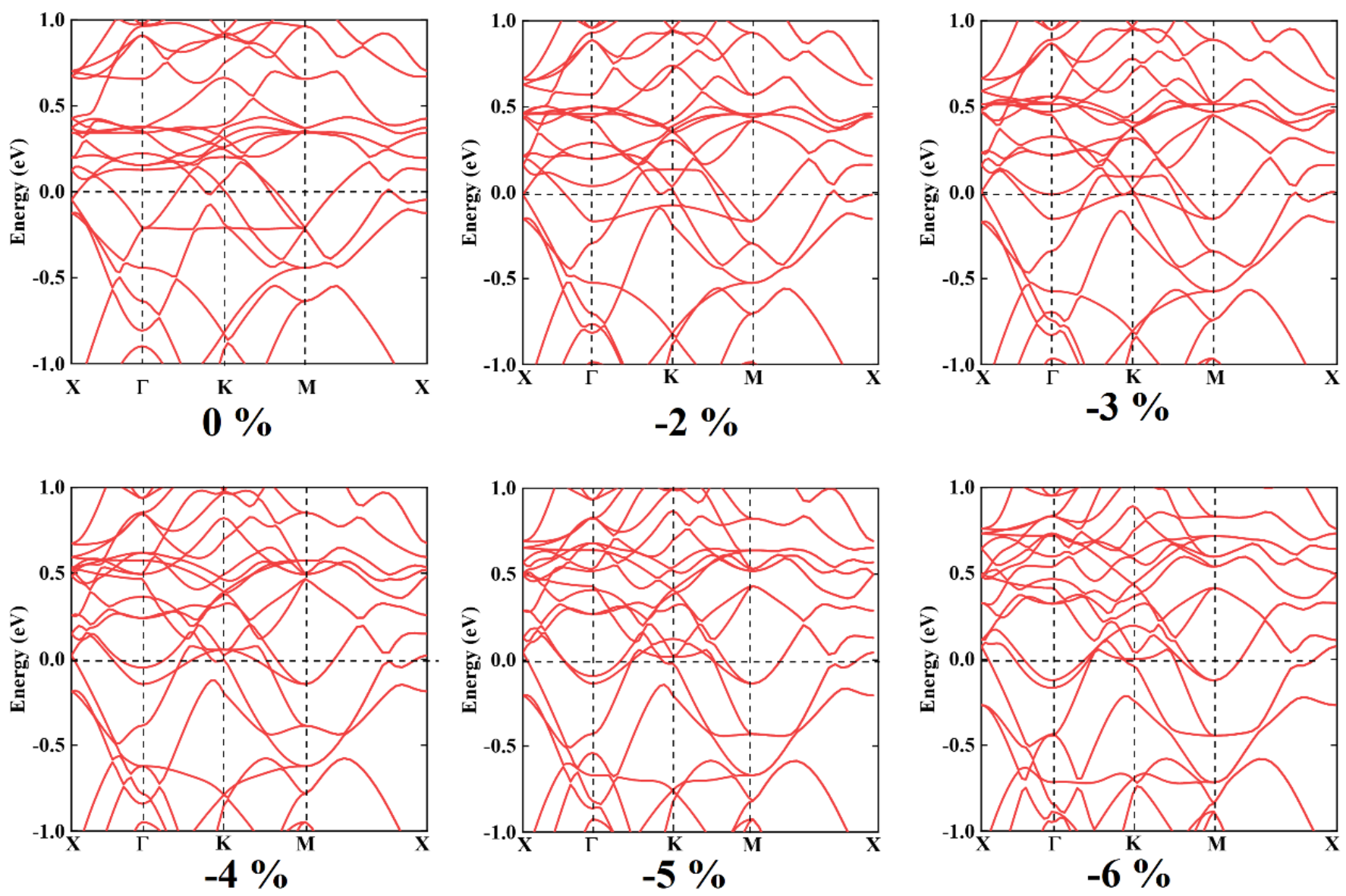

FIG. S7. Band structures of 2D $CsCr_3Sb_5$ monolayer (B-type) under various compressive strains.

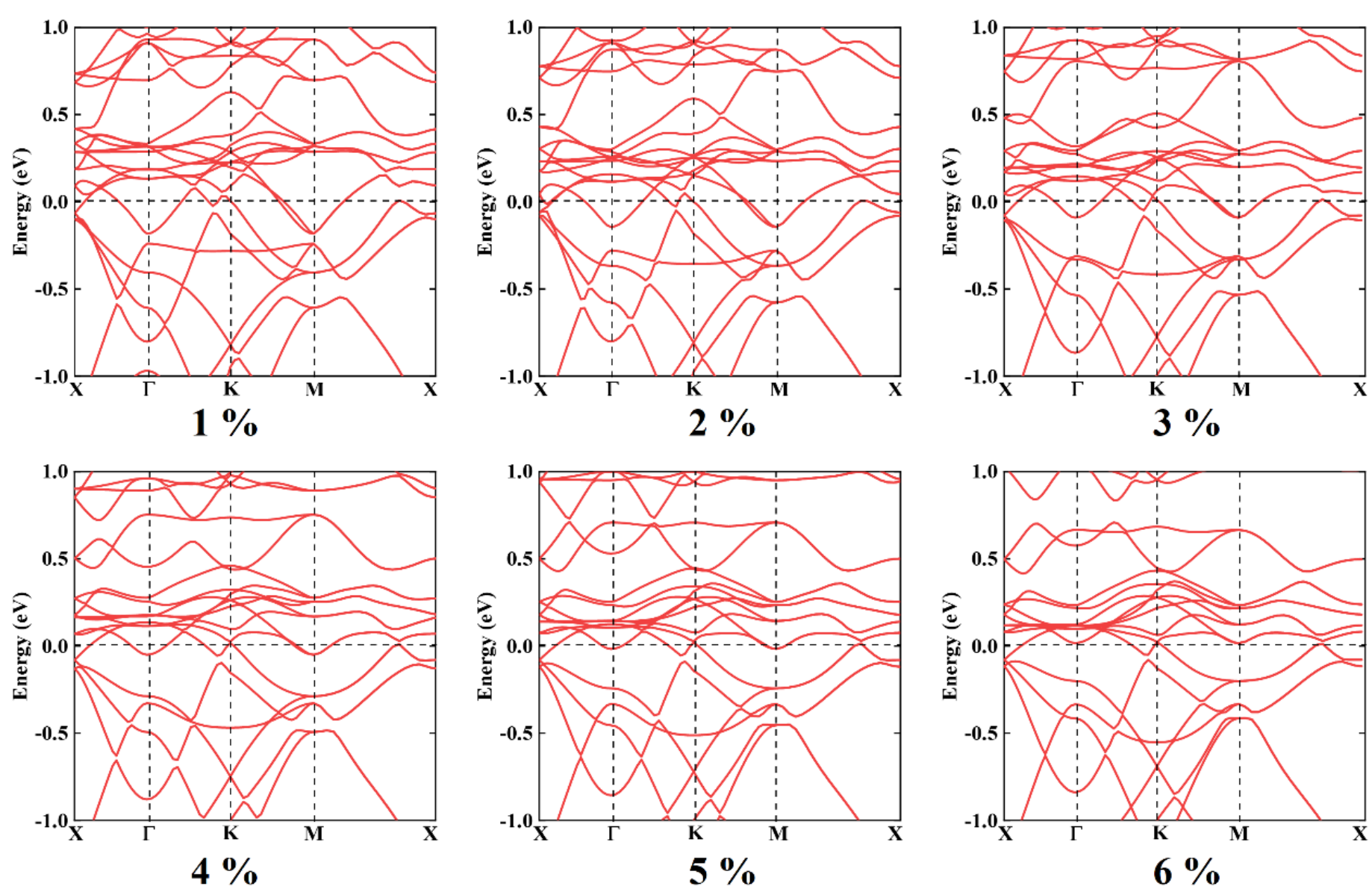

FIG. S8. Band structures of 2D $Cr_3Sb_5$ monolayer under various tensile strains.

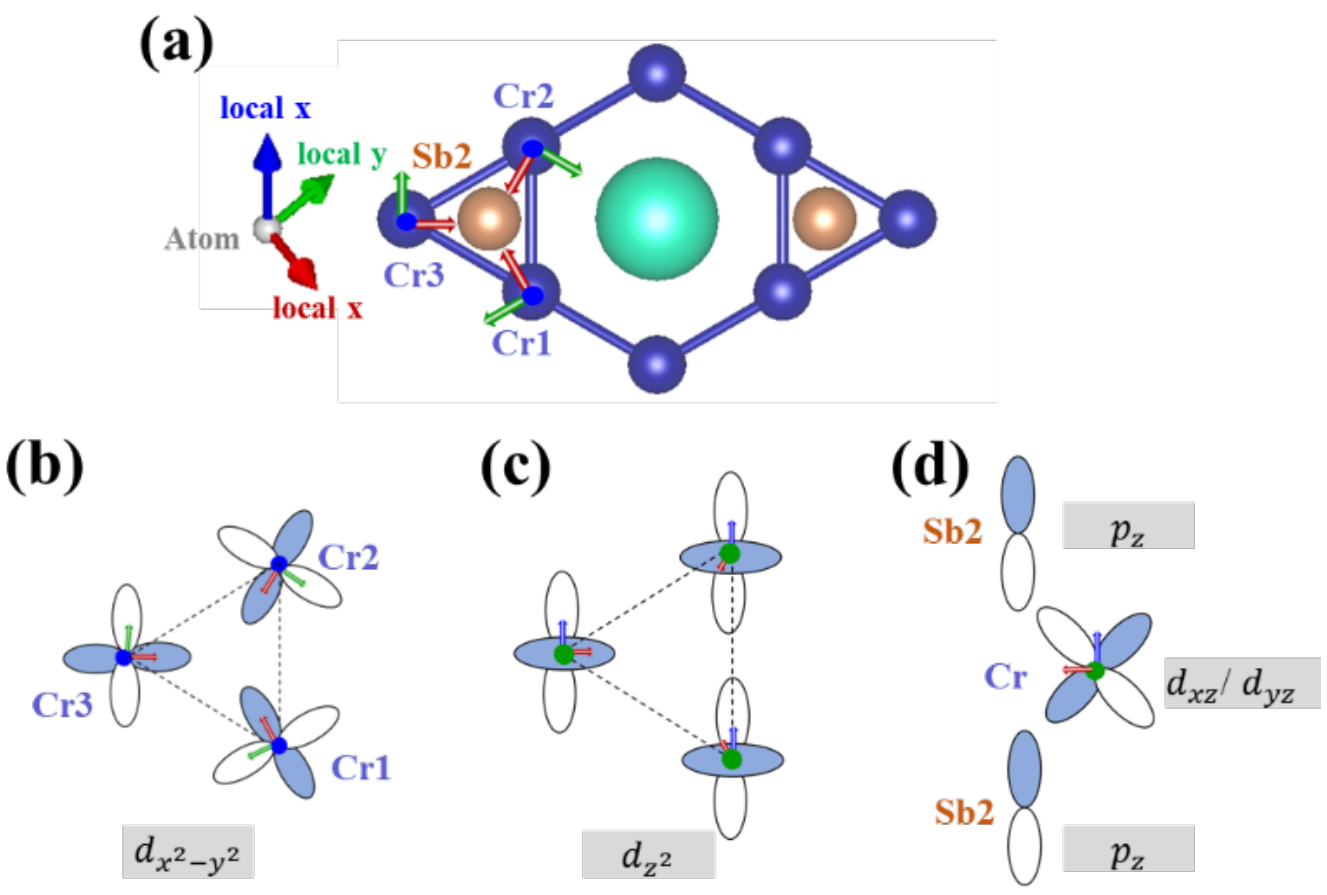

FIG. S9. (a) Local coordinates of various Cr-sublattices. (b) $d_{x^2-y^2}$, (c) $d_{z^2}$, (d) $d_{xz}/d_{yz}$ orbitals of various Cr-atoms. The hybridization of Cr $d_{xz}/d_{yz}$ orbitals with $p_z$ orbitals of Sb atoms is depicted in (d).

## C. Magnetic configurations and bands under compressive strain with SDW

Figure S10 presents different magnetic states in the two types of monolayers, and their energies are presented in table III. It is shown that the altermagnetic state (failed AF-SOD configuration) still possesses the lowest energy, suggesting that the altermagnetic state could still be the ground state in 2D $CsCr_3Sb_5$ monolayers.

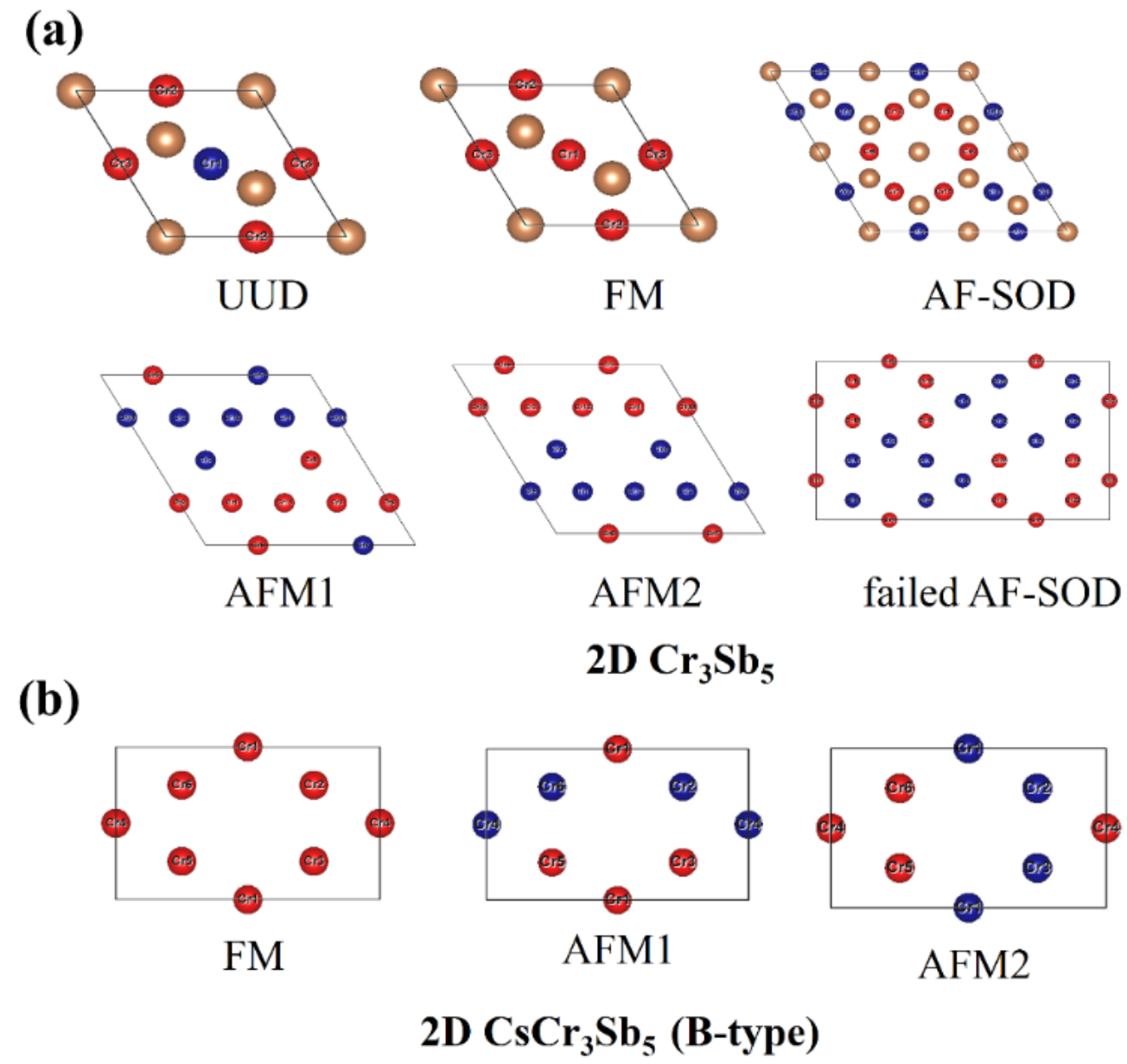

FIG. S10. (a) Spin configurations for (a) 2D $Cr_3Sb_5$ and (b) B-type $CsCr_3Sb_5$ monolayer. The failed AF-SOD configuration has also been considered in B-type $CsCr_3Sb_5$ monolayer.

Table II. Total energy (in eV), volume (in Å$^3$), total energy per unit cell (meV/f.u.) for the three types of $CsCr_3Sb_5$ monolayers with SDW order taken into account.

| Configuration | Energy (eV) | volume (Å$^3$) | Energy (meV/f.u.) |
|---|---|---|---|
| S-type | -445.02 | 1679.86 | -264.91 |
| 2D $Cr_3Sb_5$ | -425.37 | 875.25 | -486.00 |
| B-type | -443.47 | 1274.77 | -347.88 |

Fig. S11 is the band structure of the $4 \times \sqrt{3}$ S-type $CsCr_3Sb_5$ monolayer with spin density wave (SDW) order taken into account. After the SDW modulation, the S-type monolayer with a space group No. 10, P2/*m* does not satisfy the symmetry requirements for the altermagnetism. This phenomenon is mainly attributed to the absence of some Cs atoms in the Cs atomic layer, which causes a inconsistence between spin-up and spin-down sublattices. As a result, ***symmetry in bulk system is borken, and the S-type $CsCr_3Sb_5$ monolayer exhibits ferrimagnetism.

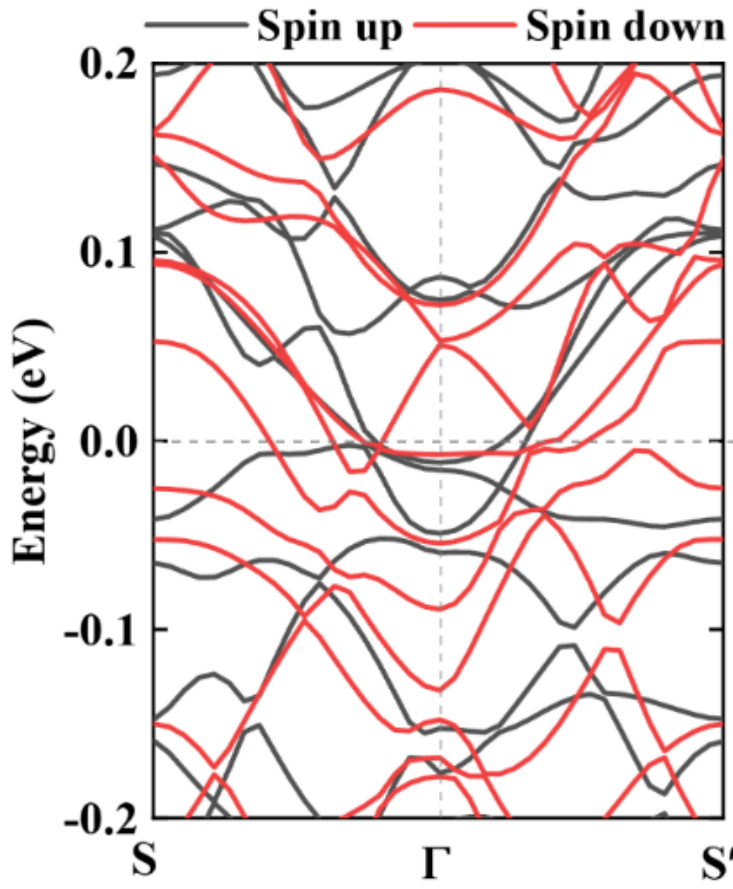

FIG. S11. Band structure of the $4 \times \sqrt{3}$ S-type $CsCr_3Sb_5$ monolayers with SDW order taken into account.

Figs. S12 and S13 present the bands and projected bands of the $4 \times \sqrt{3}$ $Cr_3Sb_5$ and B-type $CsCr_3Sb_5$ monolayers under various strains, respectively. A specific $d_{z^2}$-derived band (blue circles) undergoes a bandwidth reduction and moves to the vicinity of the $E_F$ under −4% strain, resulting in a very sharp peak in DOS. These results reveal that the electronic structures in $CsCr_3Sb_5$ monolayers are highly tunable even when the SDW is taken into account.

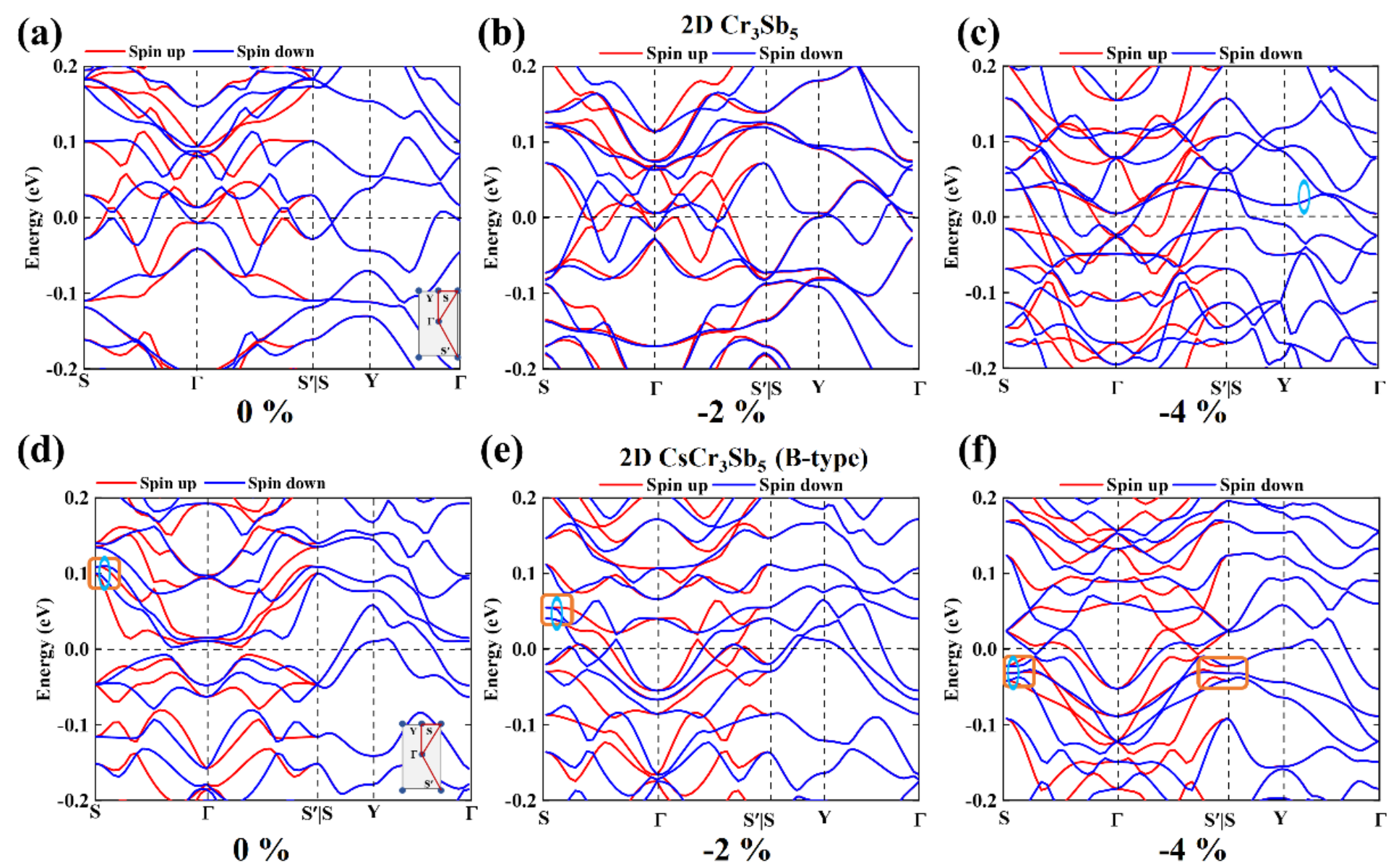

FIG. S12. Under various strains, band structures of of the $4 \times \sqrt{3}$ (a)-(c) $Cr_3Sb_5$ monolayer, and (d)-(f) B-type $CsCr_3Sb_5$ monolayer. The orange boxes and the blue circles highlight the bands that gradually shift toward the $E_F$ as the compressive strain increases.

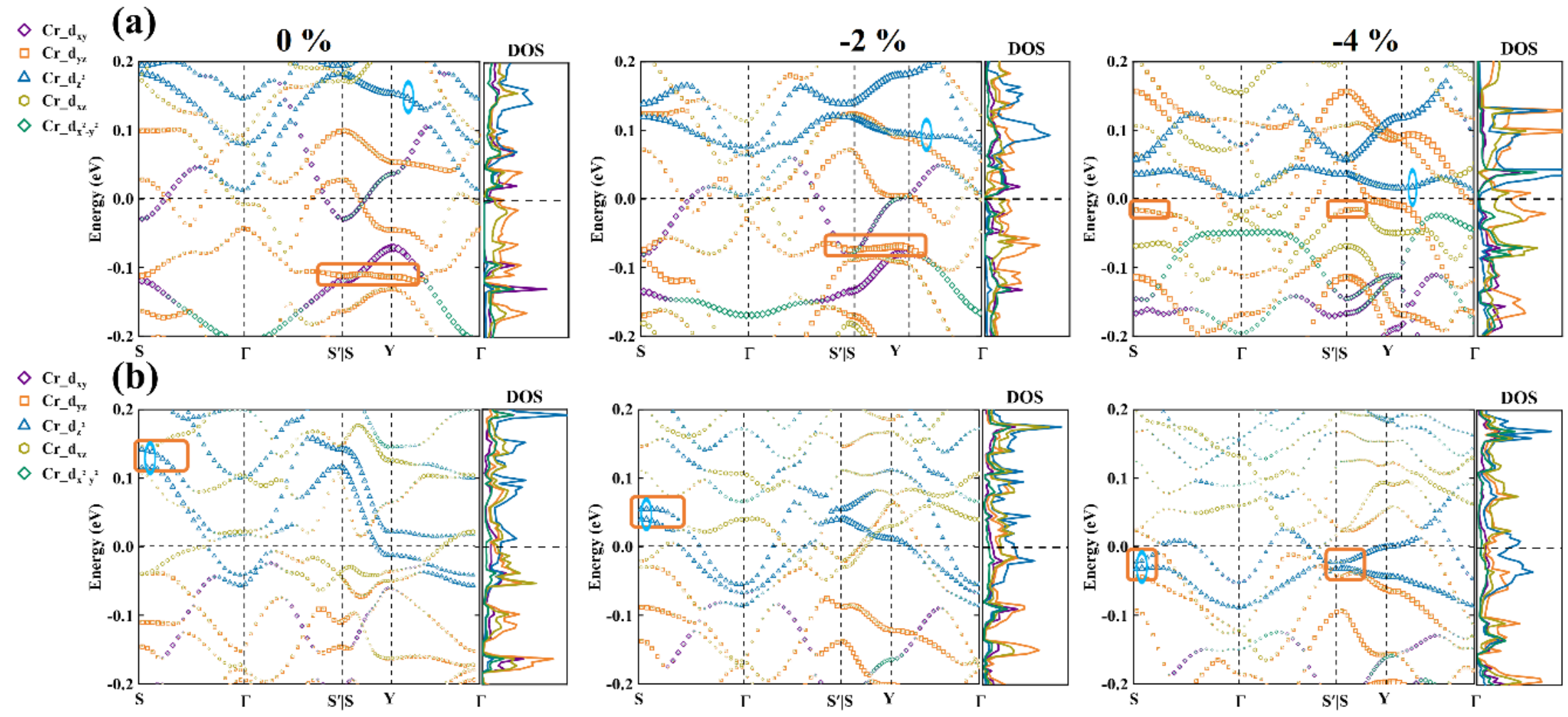

FIG. S13. Under various strains, projected bands of the $4 \times \sqrt{3}$ (a) $Cr_3Sb_5$, and (b) B-type $CsCr_3Sb_5$ monolayers. The orange boxes indicate relatively flat dispersion bands shift toward the $E_F$, giving rise to sharp peaks in DOS. The band enclosed by the blue circle narrows and gradually shifts toward the vicinity of the $E_F$ with the increasing compressive strain

Table III. Spin configurations and their energy per unit cell (meV/f.u.) for the B-types $CsCr_3Sb_5$ and $Cr_3Sb_5$ monolayers with considered SDW order.

| Structure | Configuration | Energy (eV/f.u.) |
|---|---|---|
| 2D CsCr3Sb5 (B-type) | FM | -110.53 |
| | AFM1 | -110.62 |
| | AFM2 | -110.70 |
| | failed AF-SOD | -110.87 |
| 2D $Cr_3Sb_5$ | FM | -52.98 |
| | UUD | -53.13 |
| | AF-SOD | -53.12 |
| | AFM2 | -53.11 |
| | AFM3 | -53.04 |
| | failed AF-SOD | -61.40 |